\documentclass[prl,aps,twocolumn,showpacs,preprintnumbers,amsmath,amssymb]{revtex4}
\usepackage{bm}
\usepackage{epsfig}
\renewcommand{\vec}[1]{\mathbf{#1}}

\begin{document}

\preprint{}

\title{Shubnikov de Haas effect in the metallic state of Na$_{0.3}$CoO$_2$}
\author{L. Balicas,$^1$ J. G. Analytis,$^{1,2}$ Y. J. Jo,$^1$ K.
Storr,$^{1,4}$ H. Zandbergen,$^5$ Y. Xin,$^{1}$ N. E.
Hussey,$^{3}$ F. C. Chou,$^6$ and P. A. Lee$^7$}

\affiliation{$^1$National High Magnetic Field Laboratory, Florida
State University, Tallahassee-FL 32306, USA} \affiliation{$^2$H.
H. Wills Physics Laboratory, University of Bristol, Tyndall
Avenue, Bristol BS8 1TL, UK} \affiliation{$^4$Department of
Physics, Prairie View A\&M University, Texas 77446-0519, USA}
\affiliation{$^5$Department of Nanoscience, Delft University of
Technology, Rotterdamseweg 137, 2628 AL Delft, The Netherlands}
\affiliation{$^6$Center for Materials Science and Engineering,
MIT, Cambridge, Massachusetts 02139, USA}
\affiliation{$^7$Department of Physics, Massachusetts Institute of
Technology, Cambridge, Massachusetts 02139, USA}

\date{\today}%
\begin{abstract}
Shubnikov de Haas oscillations for two well defined frequencies,
corresponding respectively to areas of 0.8 and 1.36\% of the first
Brillouin zone (FBZ),  were observed in single crystals of
Na$_{0.3}$CoO$_2$. The existence of Na superstructures in
Na$_{0.3}$CoO$_2$, coupled with this observation, suggests the
possibility that the periods are due to the reconstruction of the
large Fermi surface around the $\Gamma$ point. An alternative
interpretation in terms of the long sought-after
$\varepsilon_g^\prime$ pockets is also considered but found to be
incompatible with existing specific heat data.
\end{abstract}

 \maketitle

A number of theoretical treatments have suggested that the nature
of the superconducting pairing mechanism in hydrated Na$_x$CoO$_2$
is unconventional and that it probably corresponds to a
spin-triplet state \cite{theory,theory2,theory3,mochizuki}.
Nevertheless, the experimental situation remains unclear with heat
capacity experiments in the superconducting state suggesting that
the electronic contribution can either be described in terms of an
order parameter having nodal lines \cite{HDYang}, a hypothesis
supported by muon  spin resonance ($\mu$SR) experiments
\cite{kanigel}, or simply in terms of inhomogeneity in the Na
content \cite{cava}. Measurements of the $^{59}$Co nuclear
magnetic resonance Knight shift supports either spin-triplet
\cite{ihara} or singlet pairing \cite{kobayashi}. At the same time
$\mu$SR experiments \cite {higemoto} find no indication of static
moments in the superconducting state, implying that  time reversal
symmetry is not broken.

For conventional superconductivity, as well as for most
unconventional superconductivity scenarios, the pairing mechanism
and consequently the superconducting transition temperature,
critically depends on the electronic structure near the Fermi
level.  An accurate description of the Fermi surface (FS) is
therefore critical for the superconductivity of the hydrated
Na$_{0.3}$CoO$_2$ whose FS size and precise shape still is a
central but unsettled issue. Local-density approximation (LDA)
calculations \cite{singh} for the unhydrated NaCo$_2$O$_4$
compound indicate that two-bands, the $A_{1g}$ and one of the
$\varepsilon_{g}^{\prime}$ bands, cross the Fermi level creating
respectively, a large hexagonal Fermi surface around the $\Gamma$
point of the Brillouin zone (BZ) and six small elliptical pockets
of holes near the K point.

This Fermi surface geometry is the starting point for several of
the proposed theories of unconventionally mediated
superconductivity \cite{theory}, where the existence of small
nearly perfectly elliptical hole pockets resulting from the
$\varepsilon_{g}^{\prime}$ band is essential.
\cite{theory3,mochizuki} However, angle-resolved photoemission
(ARPES) on Na$_x$CoO$_2$ (for $0.3 \leq x \leq 0.72$) reveals only
a single FS centered around the $\Gamma$ point whose area changes
with $x$ according to the Luttinger theorem \cite{arpes, yang},
while the $\varepsilon_{g}^{\prime}$ band and the associated FS
pockets were found to sink below the Fermi energy independently of
the doping level or temperature \cite{yang}. This discrepancy
between ARPES and LDA calculations were claimed to result either
from strong electronic correlations \cite{arpes2, lee;zhang,
ishida} or Na disorder \cite{singh_again}.

For over half a century, quantum oscillatory (QO) phenomena, such
as the Shubnikov-de-Haas (SdH) bulk effect, have provided detailed
information about the geometry of the Fermi-surface of high purity
metals. Recently, we were able to observe SdH oscillations of very
small frequencies even \emph{within} the charge ordered (CO) state
of Na$_{0.5}$CoO$_2$ \cite{prlme}. The area of these orbits
$\lesssim 0.25$ \% of the hexagonal FBZ, are nearly one order of
magnitude smaller than what is expected for the pocket areas
resulting from the $\varepsilon_{g}^{\prime}$ band, see Fig. 1
(a). In Fig. 1 (b) we follow Bobroff \emph{et al.} \cite{bobroff}
and plot the orthorhombic Brillouin zone resulting from the Na
superstructure reported for $x=0.5$ \cite{zandbergen},
superimposed to the $A_{1g}$ FS assuming the absence of the
$\varepsilon_{g}^{\prime}$ pockets as suggested by ARPES. Notice
how the new BZ intersects the FS and leads to a series of smaller
pockets.

The largest of these are the oblong-shaped pockets which are well
nested by the wave vector $\vec{Q}$ show in Fig. 1(b).  Bobroff
\emph{et al.} proposed a spin density wave instability associated
with this nesting as the explanation of the magnetic ordering at
86~K. However, this ordering vector is inconsistent with that
observed by neutron scattering \cite{gasparovic}.  Furthermore,
the resistivity is largely insensitive to the magnetic transition
at 86~K \cite{gasparovic}. These observations, together with the
local moment nature of the magnetic order, suggest to us that the
explanation of the 86~K transition lies in strong correlation
physics which produces a local moment in a metallic state with
only a small degree of charge disproportionation, as seen in NMR
\cite{bobroff}.   Instead, we propose that a charge or spin
density wave with wavelength $\vec{Q}$ is responsible for the
increase in resistivity at 51~K. A similar suggestion was made by
Qian \emph{et al.} \cite{qian}. Note that the resistivity rises
sharply only below 20~K.  We interpret this rise to be due to
charge localization in the remaining two small pockets.  We
further suggest that there may be small regions in the sample
where the Na ions are particularly well ordered and localization
can be avoided.  The quantum oscillation of one or both of these
pockets may explain the SdH effect of small amplitude that we
observed in an otherwise insulating sample.  We believe that this
scenario could explain the origin of the insulating state and the
very small SdH frequencies seen by us. Interestingly, as shown in
Fig. 1(c) the incommensurate Na superstructure reported for
Na$_{0.3}$CoO$_2$ \cite{zandbergen} would also lead to a similarly
reconstructed FS although in the absence of hydration no
electronic ordering has been reported so far.

\begin{figure}
\begin{center}
\epsfig{file=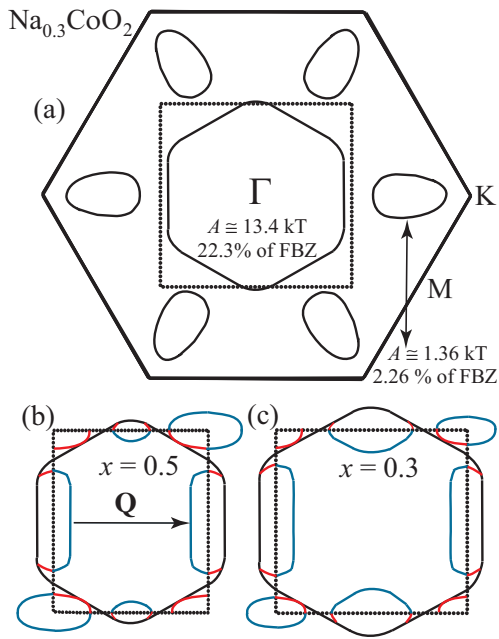, width=6.5 cm} \caption{(a)
The Fermi surface of Na$_{0.3}$CoO$_2$ within the first Brillouin
zone as calculated by the Local Density Approximation. The
corresponding Shubnikov de Haas frequencies associated with the FS
cross sectional areas are indicated. The gray rectangle depicts
the reconstructed Brillouin zone due to the Na superstructure
observed in Na$_{0.3}$CoO$_2$. (b) Assuming the absence of the
small $\epsilon^{\prime}_{g}$ pockets, the new BZ resulting from
the Na ordering seen in Na$_{0.5}$CoO$_2$ leads to a reconstructed
Fermi surface (blue and red lines). Possible nesting vectors
leading to a density-wave like instability are easily
identifiable. (c) The Na ordering observed in Na$_{0.3}$CoO$_2$
would also lead to a similarly reconstructed FS.}
\end{center}
\end{figure}

In order to explore this possibility, here we report a high
magnetic-field electrical transport study in Na$_{0.3}$CoO$_2$
single crystals at low temperatures.

\begin{figure}[htb]
\begin{center}
\epsfig{file=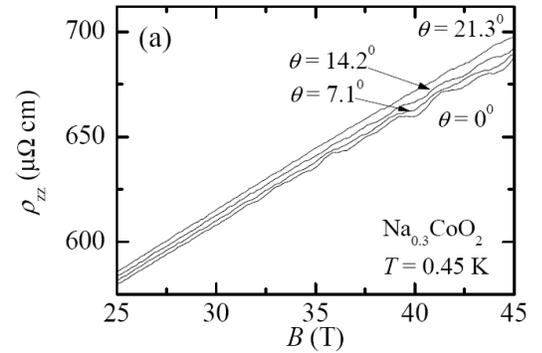, width = 6.8 cm, angle = -90}
\caption{(a) The inter-plane resistivity $\rho_{zz}$ for a
Na$_{0.3}$CoO$_2$ single crystal as a function of magnetic field
$B$ at $T \simeq 0.5$ K and for a few angles $\theta$ between $B$
and the inter-plane c-axis. Notice the presence of Shubnikov de
Haas oscillations of very small amplitude which quickly disappears
as $\theta$ increases. (b) Normalized inter-plane resistivity,
$(\rho_{zz} - \rho_{0})$/ $\rho_{0}$, where $\rho_{0} =
\rho(\theta = 0)$ as a function of $\theta$ at $H = 45$ T and $T
\simeq 0.5$ K.}
\end{center}
\end{figure}

Single crystals of Na$_{0.75}$CoO$_2$ were grown using the
floating-zone technique. By using an electrochemical
de-intercalation procedure, samples were produced with a nominal
Na concentration  $x = 0.3 \pm 0.03$, as confirmed by Electron
Microprobe Analysis. Details of the crystal growth process are
discussed in detail in Ref. \cite{chou}. Resistivity measurements
were performed in 12 single crystals using standard four-terminal
technique in a rotating sample holder inserted in a $^3$He
cryostat. Shubnikov-de-Haas oscillations were observed in only 3
crystals which showed basically the same two SdH frequencies. High
magnetic fields up to $H$ = 45 T were provided by the hybrid
magnet at the NHMFL.
\begin{figure}[htb]
\begin{center}
\epsfig{file=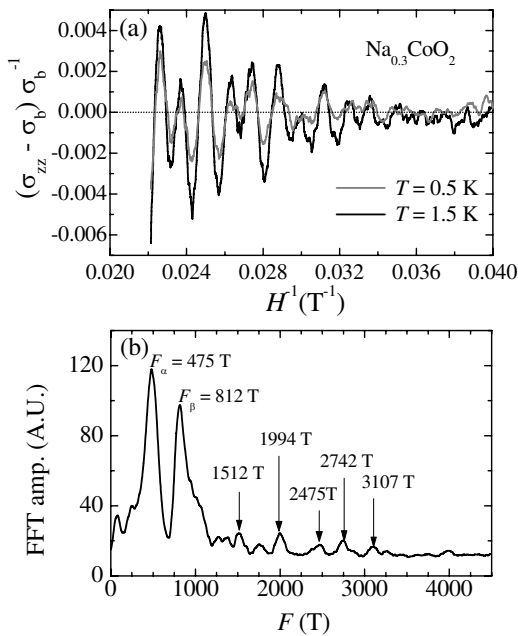, width = 6.8 cm} \caption{(a) The
Shubnikov-de Haas signal  $(\sigma - \sigma_{b})/ \sigma_{b}$,
where $\sigma = 1/\rho_{zz}$ and $\sigma_b$ is the inverse of the
background resistivity, as a function of inverse field $H^{-1}$
and for several temperatures. (b) The FFT spectrum of the SdH
signal shown in the upper panel for $T = 0.45$ K. Two pronounced
peaks at $F_{\alpha} \simeq 475$ and $F_{\beta} \simeq 812$ T are
clearly seen. Additional smaller peaks close to
  values of harmonics of $F_{\alpha}$ and $F_{\beta}$ are also
observed. }
\end{center}
\end{figure}
\begin{figure}[htb]
\begin{center}
\epsfig{file=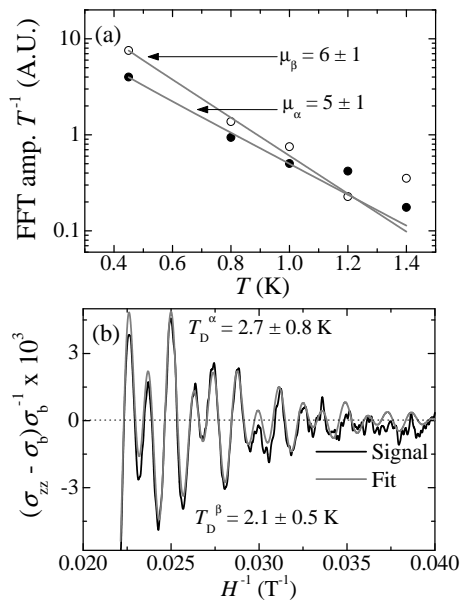, width=6 cm} \caption{(a) The
amplitude of the FFT spectrum for both frequencies $F_{\alpha}$
and $F_{\beta}$ normalized respect to the temperature $T$. Gray
lines are fits to the Lifshitz-Kosevich expression $x/sinhx$. (b)
A fit of the SdH signal to two Lifshitz-Kosevich oscillatory
components (gray line). From the fit we extracted the Dingle
temperatures $T_{D}^{\alpha, \beta}$ }
\end{center}
\end{figure}

Figure 2 shows the inter-plane resistivity $\rho_{zz}$ of a
Na$_{0.3}$CoO$_2$ single crystal as a function of the external
field $H$ at a temperature $T \simeq 0.5$ K and for several angles
$\theta$ between $H$ and the inter-plane c-axis. Notice the
appearance of small oscillations in the resistivity, i.e.,
Shubnikov-de-Haas oscillations, that disappear very quickly as
$\theta$ increases. Fig. 3 (a) displays the SdH signal defined as
$(\sigma - \sigma_b)/ \sigma_b$, where $\sigma = 1/ \rho_{zz}$ and
$\sigma_b = 1/ \rho_b$ with $\rho_b$ being the background
resistivity as function of $H^{-1}$ for two temperatures $T=0.5$
and 1.5 K, respectively. The maximum amplitude of the oscillatory
signal corresponds to only $\sim 0.3 \%$ of $\rho_{zz}$. The FFT
spectrum of this SdH signal is presented in Fig. 3 (b). Two
pronounced peaks are observed at $F_{\alpha} = 475 \pm 50$ T and
$F_{\beta} = 800 \pm 50$ T which according to the Onsager relation
$F = A(\hbar/2 \pi e)$, where $A$ is the FS cross-sectional area
perpendicular to $H$, correspond respectively to 0.8 and 1.35 \%
of the area of the undistorted hexagonal FBZ.  Unfortunately the
very limited angular range where the oscillations were observed
did not allow us to clearly define the dimensionality of these
orbits though one would expect them to follow a $1/ \cos (\theta)$
dependence associated with two-dimensional Fermi surfaces. We also
observe a series of other much smaller peaks. Given their quite
small amplitudes however, and the fact that their frequencies are
close to harmonic values of $F_\alpha$ and $F_\beta$, we only
consider here the peaks associated with $F_\alpha$ and $F_\beta$.

Figure 4 (a) displays the temperature dependence of both FFT
amplitudes normalized with respect to $T$. Gray lines correspond
to fits to the Lifshitz-Kosevich (LK) formula $x/sinhx$ with $x =
14.69 \mu T/B$ where $\mu$ corresponds to the effective mass in
relative units of the free electron mass. We obtain $\mu_\alpha =
5 \pm 1 $ and $\mu_\beta = 6 \pm 1$ for $F_\alpha$ and $F_\beta$,
respectively. These values for the effective masses are
considerably larger than what was obtained for $x = 0.5$
\cite{prlme}. Notice that if one considers the error bars both
masses have basically the same value, suggesting that they
originate from the same Fermi surface sheet. Finally, in Fig. 3(c)
we include the SdH signal for $T = 0.45 $ K and a fit of this
signal to two Lifshitz-Kosevich oscillatory components, from which
we can extract the so-called Dingle temperature $T_D= \hbar/2 \pi
k_B \tau$, where $\tau^{-1}$ is the quasiparticle scattering rate.
The SdH signal is clearly well reproduced by just two components,
justifying our previous decision of neglecting the other peaks
seen in the FFT spectrum. It yields values between 2 and 3 K for
$T_D$, lower than those obtained for $x=0.5$ \cite{prlme}.

As shown in Fig.1~(c), the observation of two frequencies suggests
the possible existence of Na superstructure(s) that redefines the
Brillouin zone and thus the geometry of the Fermi surface for
$x=0.3$. In order to explore this hypothesis electron diffraction
measurements (EDM) were performed in this single crystal,
(equipment and procedure described in Ref. \cite{zandbergen}). The
measurements reveal the existence of several Na superstructures
having periods of 0.25[110], 0.33[110], and 0.5[110] with an
occurrence ratio of 30, 10 and 20\%, respectively, with 40\% of
the sample showing no Na pattern. This clearly indicates that Na
is inhomogeneously distributed and thus the quantum oscillatory
phenomena might emerge from regions of the sample where Na is
particularly well ordered. Although EDM studies in all 3
single-crystals could not reveal a common superstructure among
them, we cannot discard this hypothesis without further work.

We now discuss a second possible interpretation of our data.  We
note that cross-sectional areas we observed are quite close to
those predicted by the LDA calculations for the cylindrical Fermi
surfaces having elliptical cross-sections resulting from the
$\varepsilon_{g}^{\prime}$ band for Na$_{0.5}$CoO$_2$, i.e., 0.6
and 1.4\% of the hexagonal FBZ, respectively \cite{singh}. We have
not been able to detect the much higher frequency associated with
the $A_{1g}$ band-derived FS observed by ARPES \cite{arpes, yang}.
According to LDA, one expects \emph{two} frequencies as a result
of the corrugation of the cylindrical Fermi surfaces from
$\varepsilon_{g}^{\prime}$ band due to a finite inter-plane
coupling. This leads to a maximal cross-sectional area within the
$\Gamma$-K plane and to a minimal one in the A-H plane of the FBZ
\cite{singh}. In fact, neutron scattering experiments
\cite{keimer} have pointed to the relevant role of the inter-plane
exchange interaction in stabilizing the three-dimensional magnetic
structures seen in Na$_x$CoO$_2$. This has been explained in terms
of superexchange interactions via, for example, O-O hopping which
would \emph{couple} each Co to its nearest inter-planar neighbors
\cite{johannes}. However, we should point out that according to
LDA the pocket area increases with decreasing $x$.  For example,
for Na$_{0.3}$CoO$_2$ the pocket shown in Ref. [16] is a factor of
2 larger in area than what we quoted earlier for
Na$_{0.5}$CoO$_2$. This discrepancy may well be within the
accuracy of LDA, especially given the possibility of strong
correlation corrections \cite{arpes2, lee;zhang, ishida}. Indeed,
Zhou \emph{et al.} \cite{arpes2} argued that strong correlation
pushes the $\varepsilon_g^\prime$ band below the Fermi energy.
However, this effect weakens for smaller $x$, and may conceivably
leave a small pocket by the time one reaches $x = 0.3$.

One difficulty with this interpretation lies in the disagreement
with ARPES data, which clearly show that the top of the
$\varepsilon_g^\prime$ band lies at 0.2~eV below the Fermi energy
at $x = 0.3$ \cite{yang}. While the reconstruction of the large
Fermi surface by Na superstructure may be beyond the resolution of
ARPES, the sinking of the $\varepsilon_g^\prime$ band is not. A
possible way out is to argue that the delicate placement of the
$\varepsilon_g^\prime$ band may be surface sensitive and ARPES is
probing only the top few layers.  However, a more serious
difficulty arises when we try to reconcile this interpretation
with specific heat data.  In two dimensions the linear coefficient
$\gamma$ of the specific heat depends only on the mass and not the
area of the pocket, with each pocket contributing $\gamma_1 = 3.4
$~mJ/Co-mole-K$^2$ for $m/m_e \simeq 5$.  For 6 pockets, this
gives $\gamma \approx 20$~mJ/Co-mole-K$^2$. Whilst this estimate
is only a factor of 2 larger than the observed $\gamma$ of
12~mJ/Co-mole-K$^2$ \cite{jin,yokoi}, the discrepancy becomes much
worse when we include the contribution of the large $A_{1g}$
pocket. ARPES measurements found that the Fermi velocity is
reduced from the LDA value by about a factor of 3 and from the
data on Na$_{0.3}$CoO$_2$ we extract $\hbar v_F \approx
0.5$~eV-\AA \cite{qian}.  Hence the $A_{1g}$ pocket alone should
contribute about 10 mJ/Co-mole-K$^2$, sufficient to account for
most, if not all of the measured $\gamma$. This puts a severe
upper limit on the mass of the $\varepsilon_g^\prime$ pockets, if
they exist. Even allowing for a factor of two uncertainty in our
measured mass, our data cannot be reconciled with existing ARPES
and specific heat data if interpreted as bulk properties due to
the $\varepsilon_g^\prime$ pockets.  Given these difficulties, we
favor the first interpretation that the SdH oscillations are due
to reconstruction of the $A_{1g}$ Fermi surface in a small part of
the sample with a particular Na superlattice.

In summary, we have presented a detailed electrical transport
study on Na$_{0.3}$CoO$_2$ revealing the existence of small Fermi
surfaces, which can be ascribed to a reconstruction of the
Brillouin zone by a particular Na superstructure. The observations
of SdH in Na$_{0.3}$CuO$_2$ and Na$_{0.5}$CuO$_2$ are important
steps towards a description of the superconducting state in
Na$_{0.3}$CoO$_2$$\cdot1.4$H$_2$O and the charge ordered state
seen in Na$_{0.5}$CoO$_2$.

We acknowledge fruitful discussions with D. J. Singh, R. L. Kurtz
and P. T. Spranger. The NHMFL is supported by NSF through
NSF-DMR-0084173 and the State of Florida. LB acknowledges the
NHMFL in-house research program. JGA acknowledges support from
Lloyd's Tercentenary Foundation and YJJ from the NHMFL-Schuller
postdoctoral program. FCC acknowledges support from the MR-SEC
Program of NSF under award number DMR-02-13282 and from DOE under
grant number DE-FG02-04ER46134. PAL acknowledges DOE number
DE-FG02-03ER46076. NEH and JA acknowledge support from the EPSRC
(UK).

\end{document}